# Growth conditions, structure and superconductivity of pure and metal-doped $FeTe_{1-x}Se_x$ single crystals

**D J Gawryluk[1], J Fink-Finowicki[1], A Wiśniewski[1], R Puźniak[1], V Domukhovski[1], R Diduszko[1,2], M Kozłowski[1,2] and M Berkowski[1]**

[1]Institute of Physics, Polish Academy of Sciences, Aleja Lotników 32/46, PL-02-668 Warsaw, Poland
[2]Tele and Radio Research Institute, Ratuszowa 11, PL-03-450 Warsaw, Poland

E-mail:gawryluk@ifpan.edu.pl

**Abstract**
Superconducting single crystals of pure $FeTe_{1-x}Se_x$ and $FeTe_{0.65}Se_{0.35}$ doped with Co, Ni, Cu, Mn, Zn, Mo, Cd, In, Pb, Hg, V, Ga, Mg, Al, Ti, Cr, Sr or Nd into Fe ions site have been grown applying Bridgman's method. It has been found that the sharpness of transition to the superconducting state in $FeTe_{1-x}Se_x$ is evidently inversely correlated with crystallographic quality of the crystals. Among all of the studied dopants only Co, Ni and Cu substitute Fe ions in $FeTe_{0.65}Se_{0.35}$ crystals. The remaining examined ions do not incorporate into the crystal structure. Nevertheless, they form inclusions together with selenium, tellurium and/or iron, what changes the chemical composition of host matrix and therefore influences $T_c$ value. Small disorder introduced into magnetic sublattice, by partial replacement of Fe ions by slight amount of nonmagnetic ions of Cu (~ 1.5 at%) or by magnetic ions of Ni (~ 2 at%) and Co (~5 at%) with spin value different than that of Fe ion, completely suppresses superconductivity in $FeTe_{1-x}Se_x$ system. This indicates that even if superconductivity is observed in the system containing magnetic ions it can not survive when the disorder in magnetic ions sublattice is introduced, most likely because of magnetic scattering of Cooper pairs.

## 1. Introduction

The simple chemical formula and simple crystallographic structure are the reasons to consider the non-stoichiometric iron selenide $FeSe_{1-\delta}$ as a model system for the investigations of mechanism of superconductivity in the iron-based superconductors. Under high pressure, critical temperature ($T_c$) for this compound increases from ~ 8.5 K up to ~ 37 K [1-4]. Quite likely, similar effect may be achieved by appropriate isovalent substitution since chemical pressure, caused by tellurium substitution into selenium site, raises critical temperature in $FeTe_{0.5}Se_{0.5}$ up to ~ 15 K [5,6]. Unfortunately, $FeTe_{0.5}Se_{0.5}$ exhibits coexistence of two tetragonal phases [7-9]. However, the materials with lower selenium content and with slightly lower transition temperature than that of $FeTe_{0.5}Se_{0.5}$ may be grown as a single phase crystals. A lot of chemical substitutions into $FeTe_{1-x}Se_x$ were already reported. Most of the work was done on polycrystalline samples and it was assumed quite often that analysed final chemical composition is identical to that one of the mixture of starting chemicals. Therefore, in order to reexamine critically very wide spectrum of possible metal substitutions, we have decided to grow single crystals of $FeTe_{1-x}Se_x$ substituted with various metals into Fe position.

Recent report by Wu *et al* [10] described the effect of various metals ion substitution, such as: Al, Ti, V, Cr, Mn, Co, Ni, Cu, Ga, In, Ba and Sm into superconducting $FeSe_{1-\delta}$, obtained by a solid state reaction. The authors claimed that indium does not substitute properly Fe and an additional InSe phase was observed. The 10 at% of Ba substituted into $FeSe_{1-\delta}$ sample destroys crystallinity. Substitution over 25 at% of Ga or Sm causes phase separation. The tetragonal structure was retained with substitution of 10 at% of the Ga and Sm, as well as with 25 at% of the aluminium. Impurities, found by X-ray analysis, were described as an additional hexagonal phase or as binary metal selenides.

All of the examined transition metals (Ti, V, Cr, Mn, Co, Ni and Cu) were found to be incorporated into the FeSe structure as far as their content does not exceed 10 at%. In previous studies, substitution of Co, Ni, Cu and Mn ion into Fe-Se system was reported as well [11-14]. It was claimed that Fe ions were successfully substituted by Co up to 50 at% [14] while the limit of solubility of Cu in the $FeSe_{1-\delta}$ reached the range 20 – 30 at% [15]. Over that, successful intercalation with 10 at% of sodium was reported too [16].

On the other hand, it was found that $FeSe_{1-\delta}$ with Fe substituted by 10 at% of Ti, V, and Cr is no longer superconducting [10]. Only the $Fe_{1-x}Cu_xSe_{1-\delta}$ samples with $x$ up to 0.02 exhibit superconductivity evidenced in transport measurements while those with $x$ over 0.03 show semiconducting behaviour [10,11]. The above results agree with those published by Williams *et al* [15] however, it was found that DC magnetic susceptibility shows no evidence of bulk superconductivity even for $Fe_{0.995}Cu_{0.015}Se$, i.e., for the sample with Cu content as low as 1.5 at%. Similar behaviour was observed for $Fe_{1-y}Ni_ySe_{1-\delta}$ system with $0.01 < y < 0.1$ [12,13]. According to Mizuguchi *et al* [13], the onset of critical temperature measured by resistivity ($T_c^{res,onset}$) was estimated to be 10 K, for the sample with $y = 0.05$. Zhang *et al* [12] reported the offset of critical temperature ($T_c^{res,offset}$) below 2 K for the sample with $y = 0.01$ and no superconducting transition above 2 K for the sample with $y = 0.04$. For $Fe_{1-z}Co_zSe_{1-\delta}$, the $T_c^{res,onset}$ was estimated to be equal to 10 K and to 5 K for the samples with $z$ equal to 0.05 and to 0.1, respectively [13]. For the sample with $z = 0.025$, the $T_c^{res,onset}$ was observed at around 2 K. It was reported that $FeSe_{1-\delta}$ samples substituted with Co concentration higher than 2.5 at% are no longer superconducting [14]. These results contrast with those of Liu *et al* [16], where $Fe_{0.92}Co_{0.08}Se_{1-\delta}$ was reported to be superconducting with $T_c$ being even higher than that of $FeSe_{1-\delta}$.

Generally, superconducting transition temperature for $FeSe_{1-\delta}$ substituted with isovalent Co, Ni, and Cu ions was found to be strongly suppressed whereas, surprisingly, substitution with Mn with concentration up to 30 at% changed the $T_c$ value only slightly. However, nonisovalent dopants like Al, Ga and Sm influence $T_c$ differently. A drop in the $T_c^{res,onset}$ at about 8.5 K and at about 6.8 K was observed for $FeSe_{1-\delta}$ doped with 10 at% of Al and with 20 at% of Ga, respectively. In contrast, substitution of 10 at% of Sm raised the $T_c^{res,onset}$ slightly to about 10.6 K, but then at the substitution of 20 at% the $T_c^{res,onset}$ dropped to 9.2 K. Interestingly, a similar drop in $T_c^{res,onset}$ to the value of 10.6 K was observed in $FeSe_{1-\delta}$ doped with 10 at% of Ba [10,11].

Recently, Kotegawa *et al* [17] reported that $^{77}$Se-NMR measurements performed on FeSe substituted with Co indicate the electron doping to the system. Since strong spin fluctuations disappear in $(Fe_{0.9}Co_{0.1})Se$, the electron doping was considered to modify the Fermi surface, resulting in the collapse of the nesting in FeSe. Kotegawa *et al* [17] suggested that this is likely the main reason of the strong suppression of the superconductivity by the Co-substitution in this compound.

Polycrystalline $FeTe_{0.5}Se_{0.5}$ samples doped with 5 at% of Mn, Co, Ni, Cu and Zn were explored by Zhang *et al* [18]. It was found that among the studied substitutions only Zn ions do not incorporate into the host lattice and most likely ZnSe formation takes place. Despite of that, both the pure and doped with Zn samples exhibited similar transition temperature to superconducting state both in magnetic and transport measurements. Pure tetragonal phases of $FeTe_{0.5}Se_{0.5}$ substituted with Cu, Ni, Co and Mn into Fe site were obtained according to powder X-ray diffraction results. It was claimed that superconducting properties are extremely sensitive to a kind of substituted metal. Superconductivity is destroyed completely in the samples with Ni and Cu substitutions in contrast to those substituted with Mn and Co, where according to the transport and magnetic measurements the $T_c$ value is altered only slightly. Shipra *et al* [19] reported successful iron substitution in $FeTe_{0.5}Se_{0.5}$ by Ni and Co up to 10 at%. They found that for both Co and Ni substituted systems strong suppression of superconducting transition temperature takes place. Successful intercalation with up to 100 at% of lithium by electrochemical technique was reported too [20].

All of the samples with various elements substituting Fe in the both systems of FeSe and $FeTe_{1-x}Se_x$ described so far were prepared by conventional solid state reaction. Only in the work of Williams *et al* [15] polycrystalline samples with partial substitution of Fe by Cu in Fe-Se system prepared by liquid phase reaction were studied.

According to phase diagram of Fe-Se system [21], it is difficult to obtain large single crystals of superconducting $FeSe_{1-\delta}$ phase using one of the techniques of the growth from the melt. The isostructural, pseudobinary Fe-Te-Se system relatively easy crystallizing from the melt is much more promising for successful growth of substituted single crystals. So far, growth of superconducting $FeTe_{1-x}Se_x$ single crystals using Bridgman's method was reported by different groups [7-9,22-32] but there is a lack of data concerning any substitution into Fe site.

In this paper, the growth of $FeTe_{1-x}Se_x$ single crystals doped with various metals is reported. Scanning electron microscopy (SEM) and energy dispersive X-ray spectroscopy (EDX) have allowed to determine unequivocally which of the dopants are incorporated into the host lattice. The influence of chemical doping at Fe site, by various elements, on superconducting transition temperature has been determined.

## 2. Experimental procedure

Superconducting single crystals of $FeTe_{1-x}Se_x$ ($x = 0.3 – 0.55$) and $FeTe_{0.65}Se_{0.35}$ doped with Co, Ni, Cu, Mn, Zn, Mo, Cd, In, Pb, Hg, V, Ga, Mg, Al, Ti, Cr, Sr and Nd have been grown using Bridgman's method. All $FeTe_{1-x}Se_x$ samples were prepared from stoichiometric quantities of iron chips (3N5), tellurium powder (4N) and selenium powder (pure). Double walled evacuated sealed quartz ampoules with starting materials were placed in a furnace with a vertical gradient of temperature equal to 1.2 – 2 $^{o}$C/mm. The samples were synthesized for 6 h at temperature 680 $^{o}$C, then temperature was risen up to 920 $^{o}$C. After melting the temperature was held for 3 h, then the samples were cooled down to 400 $^{o}$C with step of 1 – 2 $^{o}$C/h and next to 200 $^{o}$C with the rate of 60 $^{o}$C/h, and finally cooled down to room temperature with the furnace. Proper adjustment of cooling velocity and/or vertical gradient of temperature in the furnace allowed us to tune the growth velocity in the range from ~ 0.5 up to ~ 8 mm/h, permitting the growth of single crystals of various crystallographic qualities. Obtained crystals exhibited cleavage plane (001) with random orientation with respect to the growth direction. The single crystals of the highest crystallographic quality exhibited well developed (100) and (101) natural planes.

In order to obtain $FeTe_{0.65}Se_{0.35}$ single crystals substituted with various metals, the following chemicals were added: 0.5 to 20 at% of Co (metallic), NiSe (pure) and CuSe (4N), 2.5 – 20 at% of Mn pieces (5N), 1 at% of Al, Ga, In, Nd (5N), HgSe powder (5N) and 5 at% of Mg pieces (5N), Ti, V, Cr, Mo pieces (5N), Sr pieces (2N), ZnSe, CdSe powder (5N), and PbTe powder (pure).

The chemical composition of matrix and inclusions in pure and doped single crystals was checked on the cleavage plane of the crystals by field emission scanning electron microscopy (FESEM) JEOL JSM-7600F operating at 20 kV. The quantitative point analysis were performed by Oxford INCA energy dispersive X-ray spectroscopy (EDX) coupled with the SEM. Phase analysis and structural refinement of the crystals were performed at room temperature by X-Ray Powder Diffraction (XRPD) using Ni-filtered Cu K$\alpha$ radiation with a Siemens D5000 diffractometer. Precision value of the $c$ lattice constant and $\Delta\omega$ value of $\omega$ scan on 004 diffraction line were obtained in single-crystal measurements on well defined, natural cleavage (00l) plane. The diffraction patterns were analyzed by the Rietveld refinement method using DBWS-9807 program [33]. The $c$ lattice constant obtained in single-crystal measurements was used as a fixed value in powder Rietveld analysis for determination of another structural parameters, i.e. $a$, $V$ and occupation number. Data on powderized single crystals were collected in the angular range $20° < 2\theta < 100°$ with step of 0.02° and averaging time of 10 s/step. All the reflections were indexed to a tetragonal cell in the space group *P4/nmm* (No. 129) of the PbO structural type with occupation Wyckoff's 2*a* site by Fe, and 2*c* site by Se/Te. It was assumed additionally that excess of Fe ions occupy 2*c* site of structural vacancy in the Se/Te plane. The measurements of AC magnetic susceptibility (field amplitude 1 and 10 Oe, frequencies 1 and 10 kHz) for all of the samples were performed with a Physical Property Measurement System (PPMS) of Quantum Design.

## 3. Results and discussion

### *3.1. Structural analysis and superconducting transition temperature for $FeTe_{1-x}Se_x$ (0.3 < x < 0.55) single crystals*

Lattice parameters and unit cell volume for the crystals of $FeTe_{1-x}Se_x$ prepared under different conditions are limited to narrow range. No impurity phase was observed for all of the obtained crystals. For the crystals with $x = 0.35$, the average lattice constant values are equal to $a = 3.8014$ Å, $c = 6.0913$ Å, $V = 88.021$ Å$^3$, and $c/a$ ratio is close to 1.6024. Since the $c$ lattice constant strongly depends on the selenium concentration [1,7,12,23-31,33-46], as one can see in figure 1, it is easy to check if segregation of Se/Te occurs during crystallization process. We determined the value of the $c$ lattice constant at the beginning and at the end of the selected single crystals. Our results have confirmed that changes of the $c$ lattice constant along the crystal growth direction are negligible. Small inclusions of iron oxides were observed on natural cleavage planes. Initial chemical compositions in the crystallizations process compared with chemical compositions, estimated by EDX analysis, and crystallographic data for the grown $FeTe_{1-x}Se_x$ single crystals are collected in table 1. For $FeTe_{0.5}Se_{0.5}$, noticeable difference between initial and estimated by EDX chemical composition as well as broadening of 004 diffraction peak ($\Delta\omega$), connected probably with separation of phases with different Se/Te ratio, was observed [7-9]. It should be noted that only small fluctuations of the Se/Te ratio were observed for the samples with $x = 0.35$.

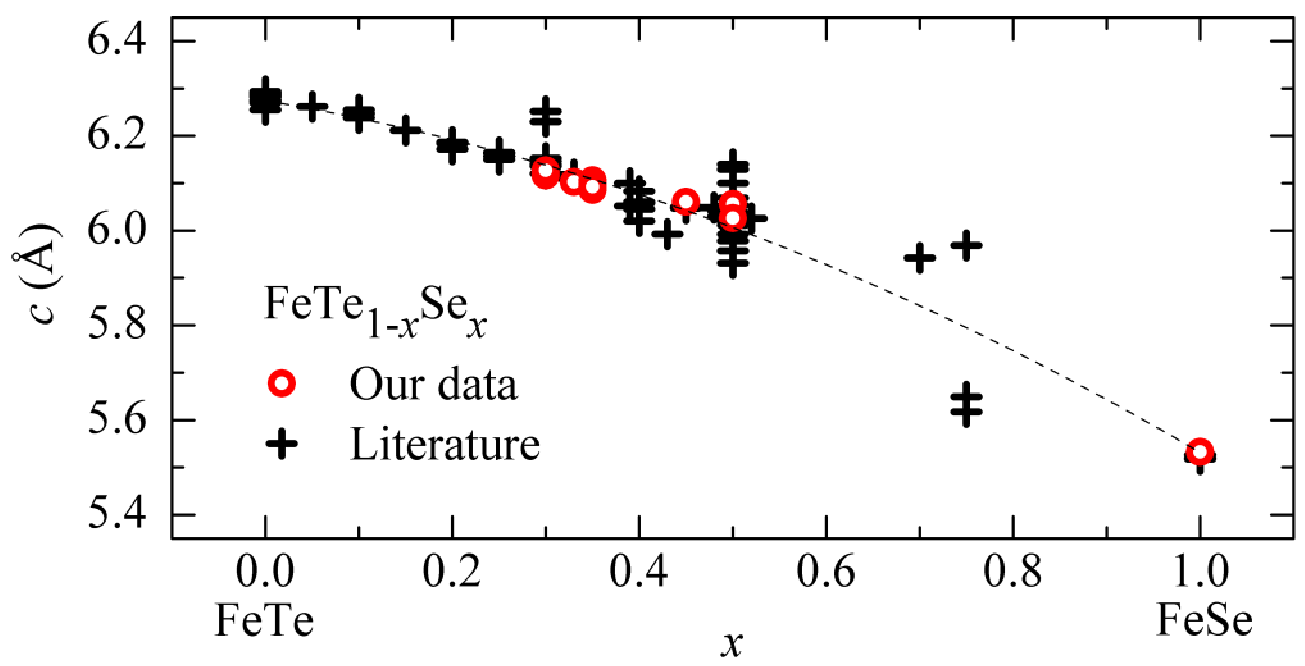

**Figure 1.** The $c$ lattice constant dependence on starting selenium content ($x$) for $FeTe_{1-x}Se_x$: comparison of our results with the literature data [1,7,12,23-31,33-46].

The critical temperature $T_c^{onset}$ of superconducting $FeTe_{1-x}Se_x$ single crystals changes from ~ 12 K for $x$ ~ 0.3 to ~ 14.7 K for $x$ ~ 0.5, in agreement with the results of previous report [34] (see, figure 2). The correction for demagnetizing field was not taken into account in the analysis of the data for temperature dependence of AC susceptibility and therefore the absolute value of presented real part of AC susceptibility is higher than 1.

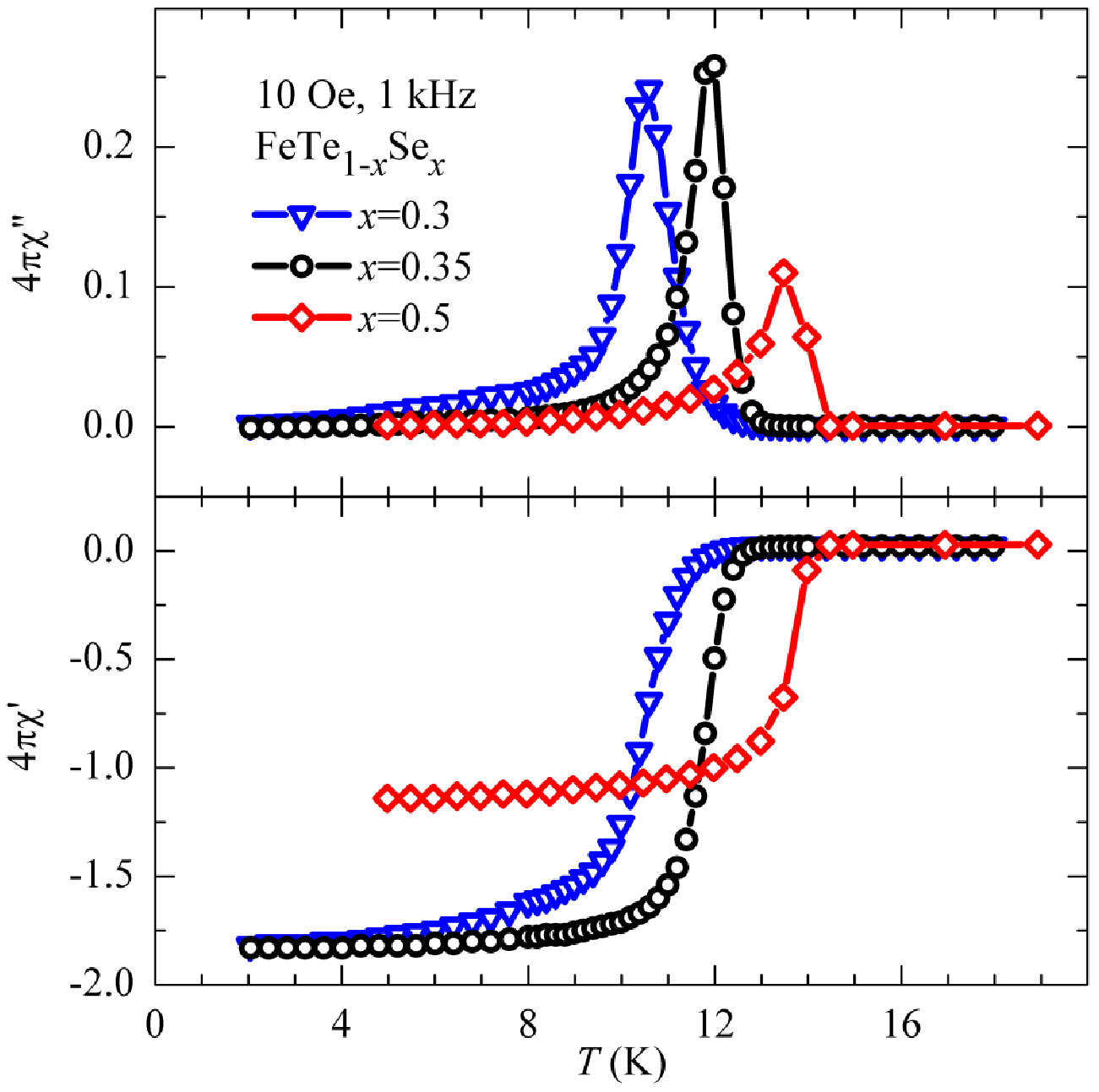

**Figure 2.** Real part (4πχ' - lower panel) and imaginary part (4πχ" - upper panel) of AC magnetic susceptibility, as a function of temperature for selected single crystals of $FeTe_{1-x}Se_x$ ($x$ = 0.3, 0.35, 0.5), measured in 10 Oe AC field with 1 kHz in warming mode. Demagnetizing field correction was not taken into account in the analysis of the data presented in the plot.

For selenium concentration $x = 0.35$, we obtained high quality single-phase crystals with critical temperature $T_c^{onset}$ of about 12.5 K. Full width at half maximum (FWHM) of 004 diffraction peak for the highest crystallographic quality single crystals of $FeTe_{0.65}Se_{0.35}$ has been found to be as small as $\Delta\omega = 1.35$ arc min. It is found that the sharpness of transition to the superconducting state is strongly correlated with the crystallographic quality as it is shown in figure 3. The crystals characterized with $\Delta\omega$ values equal to 1.67, 2.52, 3.28, and 6.00 arc min have been grown with the velocities of ~ 1.2, ~ 1.8, ~ 5, and ~ 8 mm/h, respectively. Since the investigated samples were different in shape the impact of demagnetizing field on AC susceptibility varied among them as one can see in the figure 2. Therefore, for comparison of the sharpness of superconducting state transition, the data of 4πχ' presented in figure 3 were normalized to the value of -1 for real part of AC susceptibility at low temperatures.

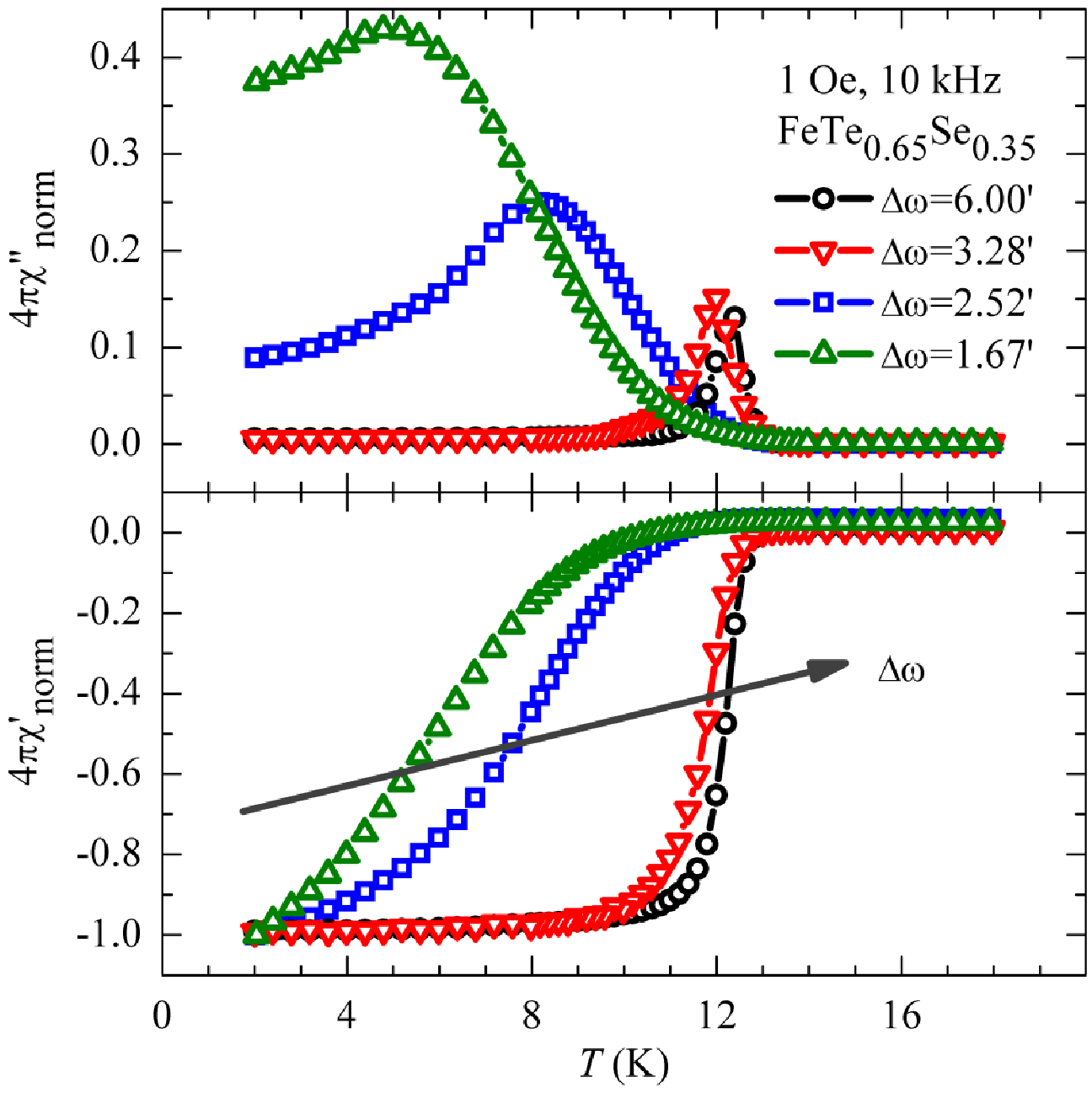

**Figure 3.** Temperature dependence of real part (lower panel) and imaginary part (upper panel) of AC magnetic susceptibility, normalized to the ideal value of –1 for real part of AC susceptibility, measured in 1 Oe of AC field with 10 kHz in warming mode for selected $FeTe_{0.65}Se_{0.35}$ single crystals with various values of Δω (listed in table 1).

It was found that the narrowest transition to the superconducting state (width ~ 0.6 K) exhibit single crystals with relatively large value of Δω equal to 6 arc min. Furthermore, the decrease of Δω value is correlated with the increase of the width of transition (90% – 10% criterion). This correlation found for almost all of the studied single crystals is not fully understood at present. It suggests that disorder is a necessary ingredient of superconducting state in $FeTe_{1-x}Se_x$ system and in any case it indicates at least that properties of $FeTe_{1-x}Se_x$ are very sensitive to the defects present in the sample [35,47,48]. It means that existence of defects in single crystal may support superconductivity.

### *3.2. Chemical substitution at Fe site by: Co, Ni, Cu, Mn, Zn, In, Pb, Hg, Cd, Mo, Mg, V, Ga, Al, Ti, Cr, Sr, and Nd in $FeTe_{0.65}Se_{0.35}$ single crystals*

The composition $FeTe_{0.65}Se_{0.35}$ was chosen to study the effect of chemical substitutions into Fe site in $FeTe_{1-x}Se_x$ system and single crystals with Co, Ni, Cu, Mn, Zn, Mo, Cd, In, Pb, Hg, V, Ga, Mg, Al, Ti, Cr, Sr or Nd dopants were grown using Bridgman's method. We have found that only the Co, Ni and Cu ions are properly incorporated into the host matrix. Substitution of Ni and Cu up to 20 at% does not lead to the appearance of any visible indication of phase separation and does not damage crystallinity. However, cobalt-iron-telluride inclusions are found in crystals with higher Co concentrations. The analysis of the chemical composition of the matrix allowed us to conclude that above 10 at% of Co competitive reaction Co-Te takes place and the limit of solubility of Co ions in the $FeTe_{0.65}Se_{0.35}$ crystals is equal to about 15 at%. SEM image of cleavage plane of the matrix of Fe-*TM*-Te-Se, where *TM* is transition metal, indicates the appearance of small iron oxides inclusions, as it is presented in figure 4.

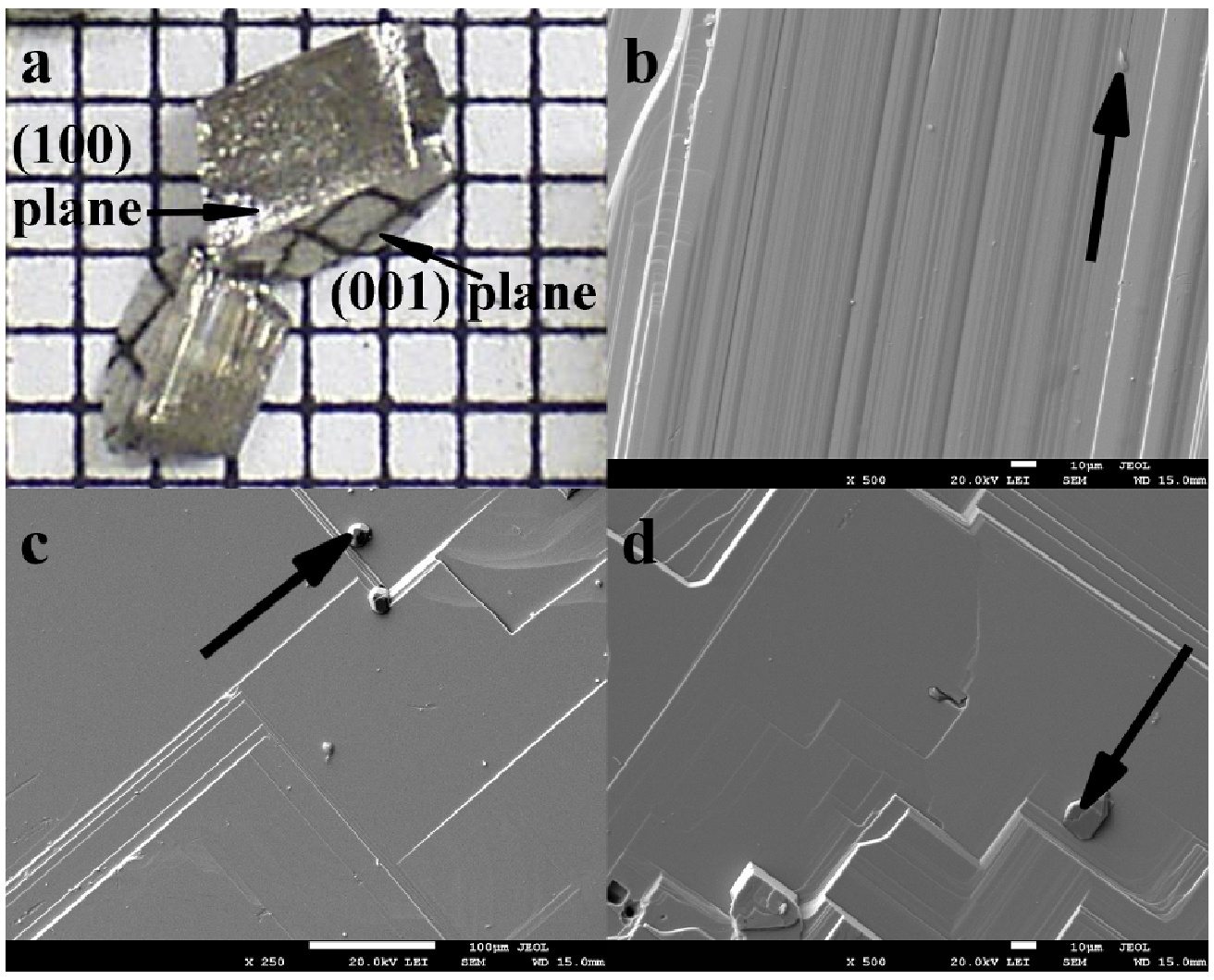

**Figure 4.** Photograph of $Fe_{0.995}Co_{0.005}Te_{0.65}Se_{0.35}$ ($\Delta\omega$ = 1.92 arc min) single crystal with visible natural (100) and (001) planes marked by arrows (a) and SEM images of the (100) crystal plane (b) compared with SEM images of natural cleavage (001) plane for the crystals with different substituent metal ion concentration and with different Te/Se ratio: $Fe_{0.995}Ni_{0.005}Te_{0.65}Se_{0.35}$ ($\Delta\omega$ = 2.45 arc min) (c) and $Fe_{0.9}Cu_{0.1}Te_{0.75}Se_{0.25}$ ($\Delta\omega$ = 2.93 arc min) (d). Arrows on the SEM images mark iron oxides inclusions. White spacers below images indicate distance of 10 μm (b, d) and 100 μm (c).

Tetragonal structure of obtained single crystals substituted with Co, Ni and Cu was confirmed by room-temperature X-ray data analysis. Full width at half maximum (FWHM) of ω scan on 004 peak for the highest crystallographic quality single crystals substituted with Ni is found to be $\Delta\omega$ = 1.25 arc min (figure 5). The data presented in the panel a) were obtained with both $K\alpha_1$ and $K\alpha_2$ irradiation and therefore two maxima are visible.

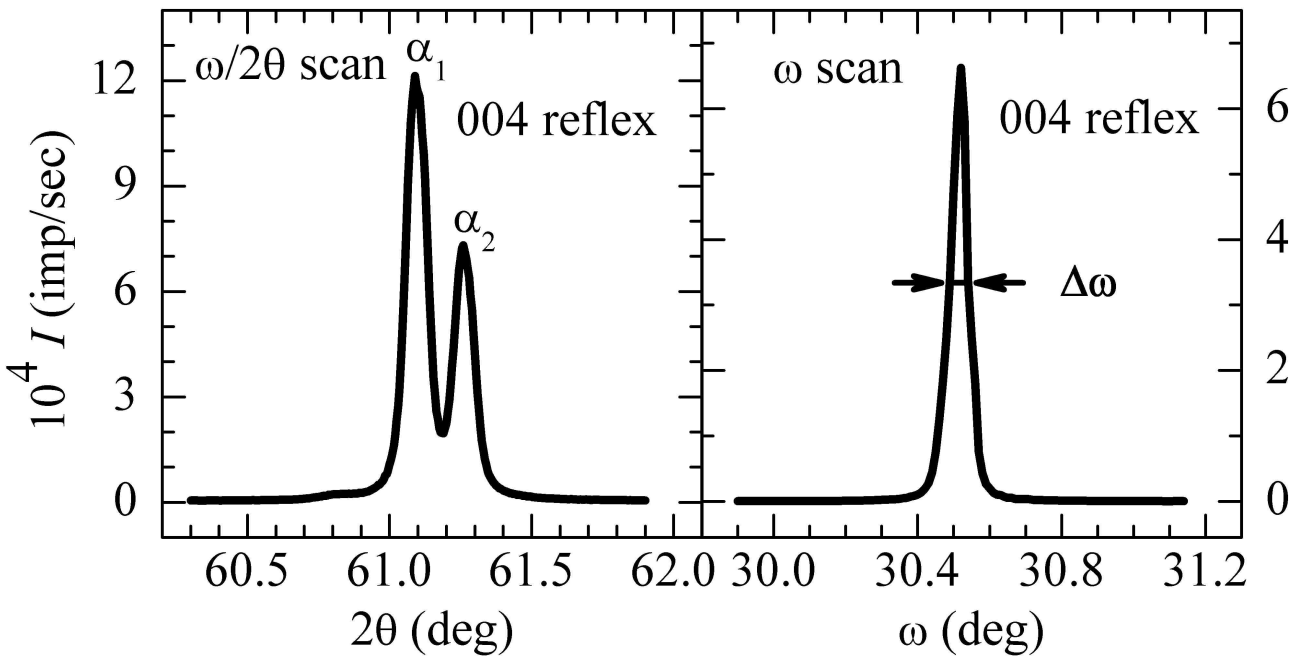

**Figure 5.** Room-temperature X-ray spectra from single-crystal measurements on natural cleavage (00l) plane of $Fe_{0.95}Ni_{0.05}Te_{0.65}Se_{0.35}$: doublet ($K_{\alpha1}$-$K_{\alpha2}$) of 004 Bragg peak obtained in ω/2θ scan (left panel) and rocking curve in ω scan measurements of 004 peak, with refined value of the full width at half maximum (FWHM ≡ $\Delta\omega$) equal to 1.25 arc min. (right panel).

It was found that the value of the *c* lattice constant decreases linearly with increasing Co, Ni and Cu ion concentration, as it is shown in figure 6. The most significant changes in the *c* value are due to copper ion substitution into Fe site while the impact of cobalt substitution is the smallest one.

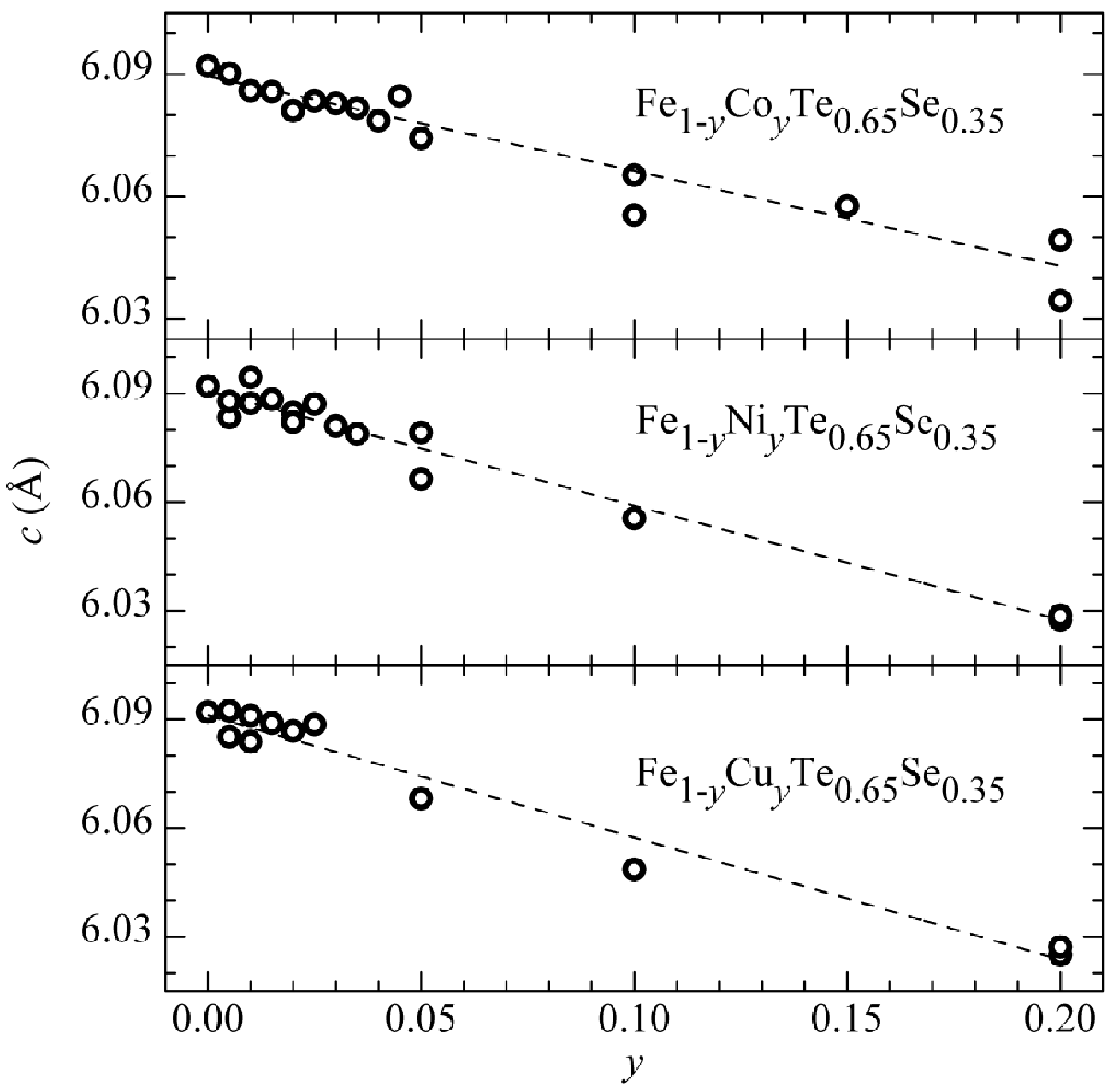

**Figure 6.** The $c$ lattice constants of $Fe_{1-y}TM_yTe_{0.65}Se_{0.35}$ (*TM* = Co, Ni and Cu) as a function of dopant content *y*. The *y* value is equal to the initial dopant concentration.

Negligible segregation of Se/Te in Fe-Te-Se system allowed us to check if the segregation between Fe and Ni or Cu takes place. We have compared the value of the $c$ lattice constant at the beginning and at the end of the grown single crystals substituted with 20 at% of Ni and Cu. The obtained results confirm (see table 2) that the changes of the $c$ lattice constant along the crystal growth direction are negligible, what means that the segregation is negligible. There is a difference between the values of lattice constant at the beginning and at the end of the crystal substituted with 20 at% of Co since 20 at% of Co is above the limit of Co solubility in Fe-Te-Se system.

Substitution of manganese and zinc leads to a creation of inclusions of metal selenides and to a change of Se/Te ratio in the host lattice, as found out by SEM/EDX. Substitution of manganese leads to a formation of an additional compound of $(MnFe)_2O_3$ type, as identified by SEM/EDX analysis. X-ray analysis indicates the presence of cubic MnSe phase (space group *Fm3m*) with average lattice constant of 5.470 Å despite that MnSe inclusions were not detected with SEM. The appearance of MnSe phase in single crystals is in an agreement with the changes of matrix composition determined from SEM/EDX analysis. Substitution of cadmium and indium leads to a creation of inclusions of CdTe or InTe type, whereas substitution of molybdenum leads to a creation of $MoTe_{1-z}Se_z$ compounds. Substitution of lead and mercury leads to a formation of metal iron telluride alloys. Substitution of vanadium and gallium leads to a formation of alloys with iron. As a result of an appearance of such iron-reach alloys, concentration of iron in the matrix is decreased. Magnesium reacts with quartz of ampoule wall and magnesium silicate compounds appear. Typical SEM images of $FeTe_{0.65}Se_{0.35}$ matrix and inclusions of ZnSe, $Mo(Te_{0.35}Se_{0.65})_2$, $V_{0.65}Fe_{0.35}$, and $Ga_{0.6}Fe_{0.4}$ are shown in figure 7.

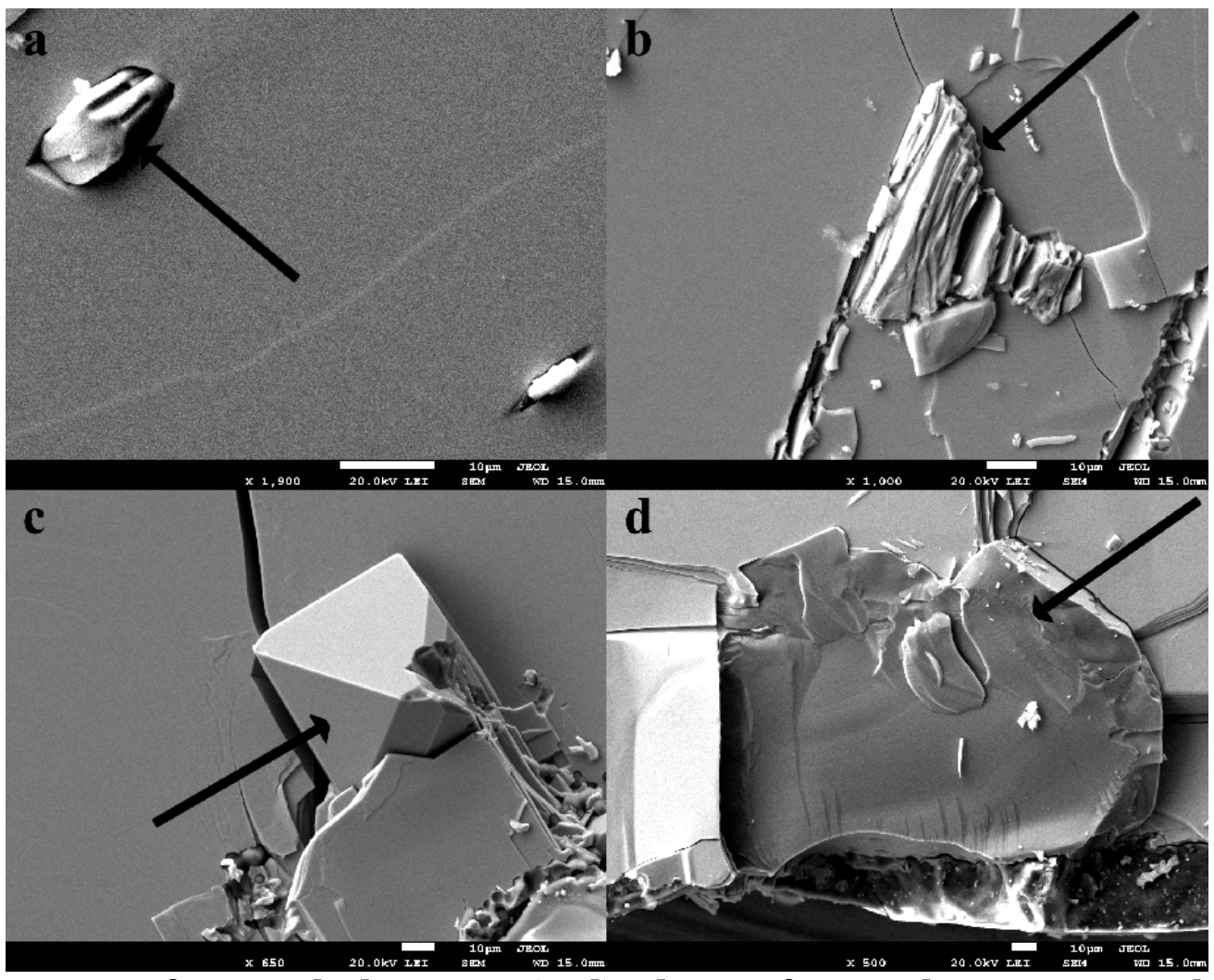

**Figure 7.** Selected SEM images of natural cleavage (00l) plane of crystal matrixes with well developed inclusions (marked by arrows) of ZnSe (a), $Mo(Te_{0.35}Se_{0.65})_2$ (b), $V_{0.65}Fe_{0.35}$ (c) and $Ga_{0.6}Fe_{0.4}$ (d) of $FeTe_{0.65}Se_{0.35}$ single crystals doped with the elements not incorporated into the host lattice. White spacers below images indicate distance of 10 μm.

It was found that the remaining added elements, i.e., Al, Ti, Cr, Sr and Nd do not incorporate into the host lattice of $FeTe_{0.65}Se_{0.35}$ single crystals. Non-reacted inclusions with Al, Ti, Cr, Sr and Nd were found only at the end of the ingots. It means that composition of the matrix is not changed significantly during crystallization. Summary of chemical compositions of the matrix and inclusions in $FeTe_{0.65}Se_{0.35}$ single crystals doped with the elements that do not incorporate into the host lattice is shown in table 3.

### *3.3. The influence of chemical substitution on the superconducting transition temperature of $FeTe_{0.65}Se_{0.35}$ single crystals*

The effect of chemical substitution at Fe site on superconducting transition temperature has been investigated and temperature dependence of AC susceptibility has been measured for all of the studied crystals. Since the experimental set-up for crystallization is a closed system all of the inclusions may affect the final product of the crystallization. Particularly, the change of Se/Te ratio or iron content in the host lattice is expected for the dopants not incorporated into the $FeTe_{1-x}Se_x$ matrix when selenides, tellurides, or binary metal compounds are formed. The concentration changes in the matrix are proportional to the amount of dopant added. When the addition of dopants leads to a creation of metal selenide compound, the selenium content $x$ in the matrix decreases. In such a case decrease of $T_c$ is expected because the relation between Se/Te ratio in the crystal and its $T_c$. We have found that it takes place in the samples doped with Mn and Mo. Nevertheless, the change of $T_c$ in the samples doped with Cd is negligible. For the samples doped with Ga and Hg, the slight increase of $T_c$ values was observed but these changes can not be related with the changes in the ratio of Se/Te in the matrix (see, table 3 with EDX data). This effect is not fully understood at the present and needs additional studies.

The decrease of $T_c$ with increasing Mn content for the samples doped with manganese is well visible in the magnetic susceptibility data (see figure 8).

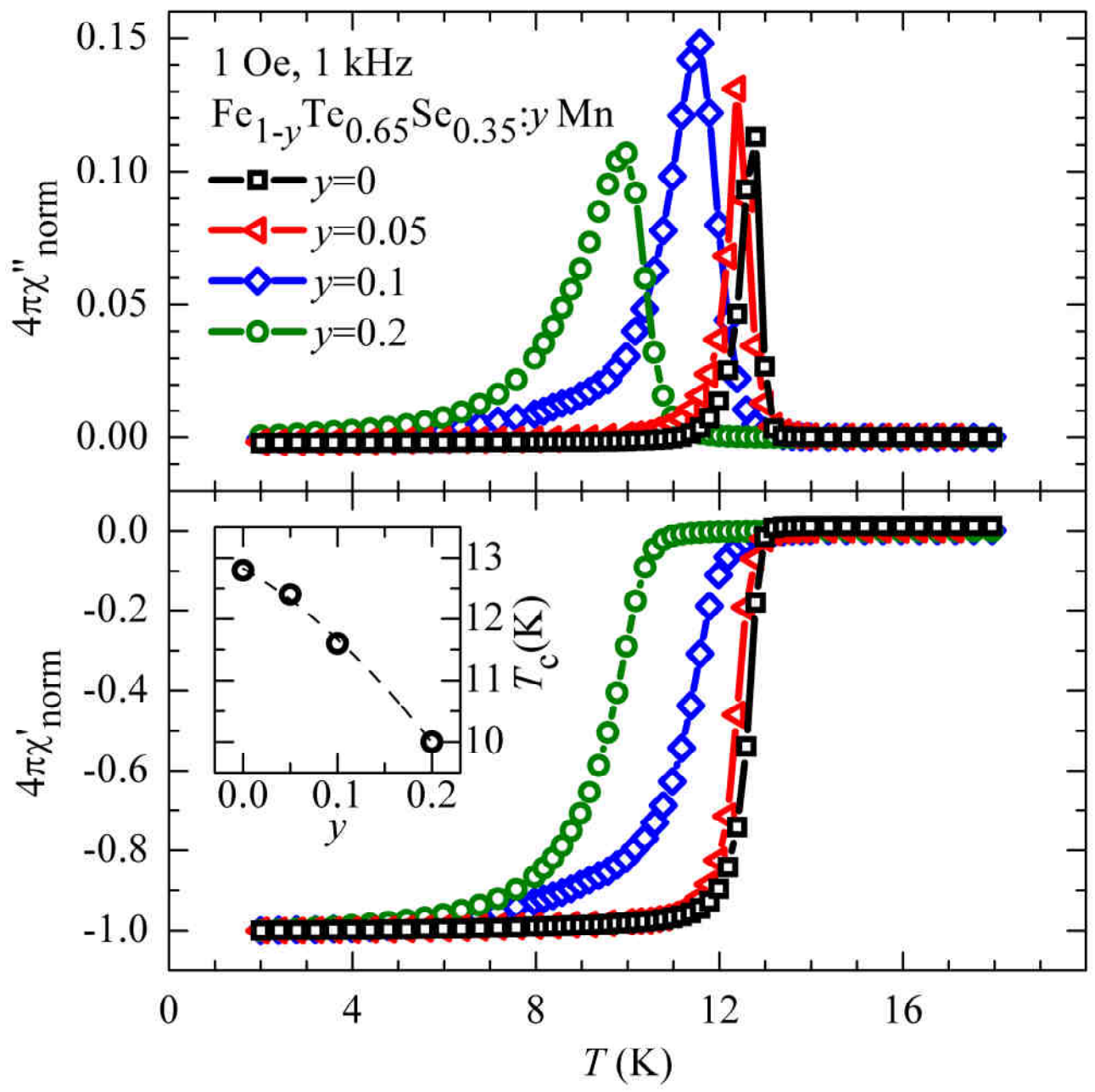

**Figure 8.** Temperature dependence of real part (lower panel) and imaginary part (upper panel) of AC magnetic susceptibility measured in 1 Oe of AC field with 1 kHz (normalized to the ideal value of –1 for real part of AC susceptibility at low temperature) for $FeTe_{0.65}Se_{0.35}$ samples doped with 5 – 20 at% of Mn instead of Fe. Manganese ions are not incorporated to the single crystals of $FeTe_{1-x}Se_x$ host lattice. The change of transition temperature is caused by decrease of selenium content $x$ in the $FeTe_{1-x}Se_x$ matrix (see comparison with figure 2). Inset shows influence of Mn concentration on critical temperature.

The observed changes in $T_c$ are caused by a decrease of selenium content $x$ in the $FeTe_{1-x}Se_x$ matrix as estimated from EDX or X-ray measurements. It was found for the studied samples that the volume of non superconducting phase increases with increasing dopant inclusion content since weakening of the intensity of diamagnetic signal was observed. The width of the transition to the superconducting state is correlated with crystallographic quality of the crystal in the same way as it was found for $FeTe_{1-x}Se_x$.

For the samples doped with cobalt, it was found that superconductivity is suppressed by 5 at% Co substitution into Fe site in single crystals of $FeTe_{0.65}Se_{0.35}$ as well as of $FeTe_{0.5}Se_{0.5}$ (composition determined by EDX expressed as $Fe_{0.9}Co_{0.06}Te_{0.53}Se_{0.47}$ with $\Delta\omega = 0.95$ arc min). All of the single crystals substituted over 5 at% of Co, 2 at% of Ni or Cu do not show superconductivity. Nevertheless, some traces of superconductivity are visible in AC susceptibility most likely due to local inhomogeneity in Co distribution. In contrast, single crystals of $Fe_{0.955}Co_{0.045}Te_{0.65}Se_{0.35}$, $Fe_{0.985}Ni_{0.015}Te_{0.65}Se_{0.35}$ and $Fe_{0.99}Cu_{0.01}Te_{0.65}Se_{0.35}$ and those with lower concentrations of Co, Ni and Cu show superconductivity. Copper substitution is the most effective in the suppression of superconductivity while the impact of cobalt substitution is the smallest one, in good agreement with theoretical calculations [49,50]. Temperature dependences of real and imaginary part of AC magnetic susceptibility for $Fe_{1-y}TM_yTe_{0.65}Se_{0.35}$ ($TM$ = Co, Ni and Cu $0.005 < y < 0.2$) crystals are shown in figures 9, 10, 11 and 12. For all of the studied samples, critical temperature and superconducting volume fraction decrease with increasing dopant content. It seems that recently described effect of iron excess on superconducting properties of pure Fe-Te-Se system [23,28,34,38,50-54] is much smaller than the impact of the *TM* dopant, especially on the critical temperature value. However, taking into account that there is no simple relation between Fe/*TM* (*TM* = Co, Ni or Cu) ratio sufficient to alter the compound from superconducting to nonsuperconducting one it is not trivial at all to make any prediction about necessary conditions for the substituted system to maintain superconductivity (for example, compare two samples substituted with 1% of Cu, see table 2).

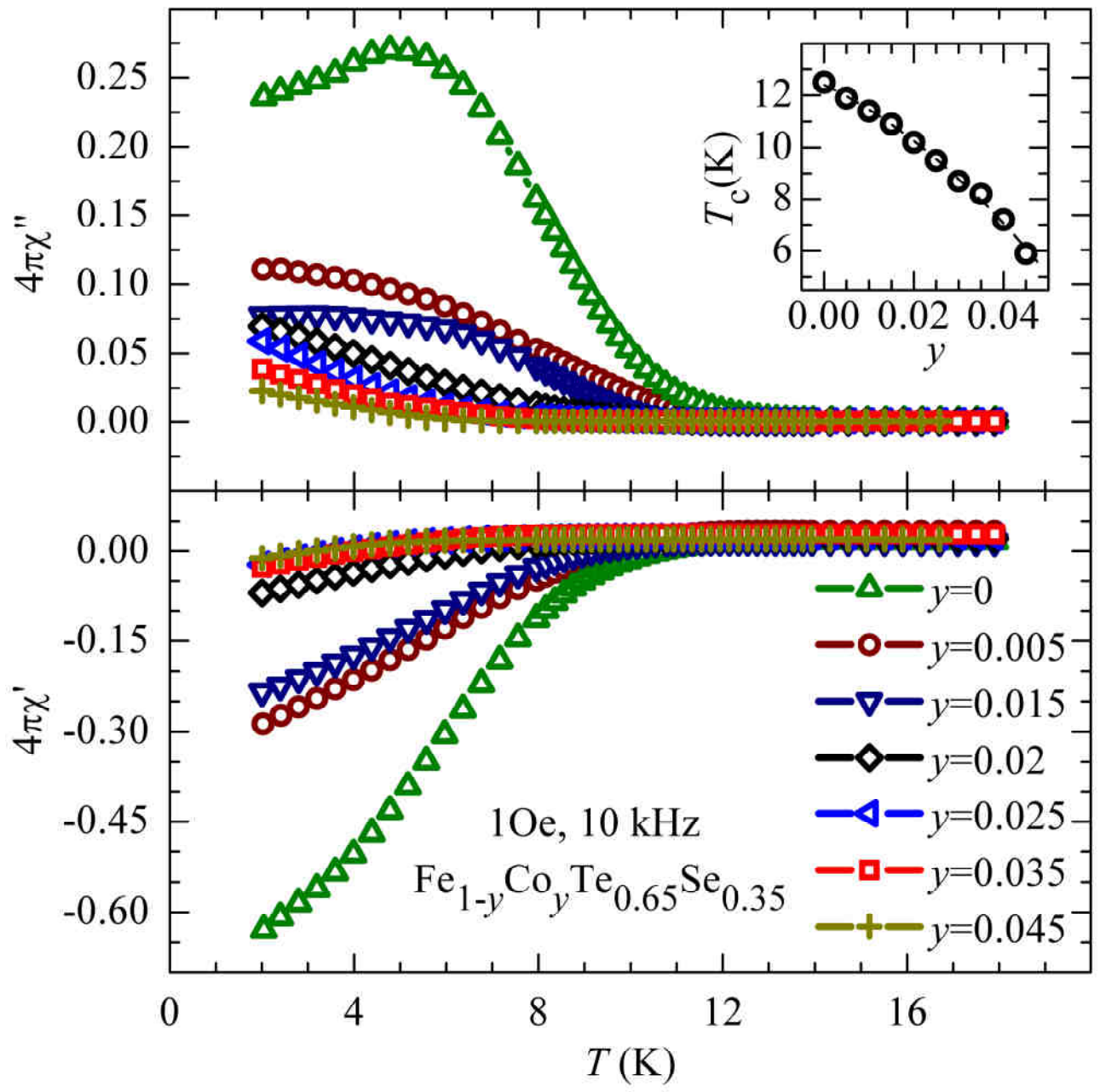

**Figure 9.** Temperature dependence of real part (lower panel) and imaginary part (upper panel) of AC susceptibility for high crystallographic quality $Fe_{1-y}Co_yTe_{0.65}Se_{0.35}$ ($0.005 < y < 0.045$) single crystals (see, table 2 for Δω) compared with that one for similar crystallographic quality $FeTe_{0.65}Se_{0.35}$ crystals (Δω = 1.67 arc min), measured in 1 Oe of AC field with 10 kHz. Inset shows critical temperature dependence on Co content.

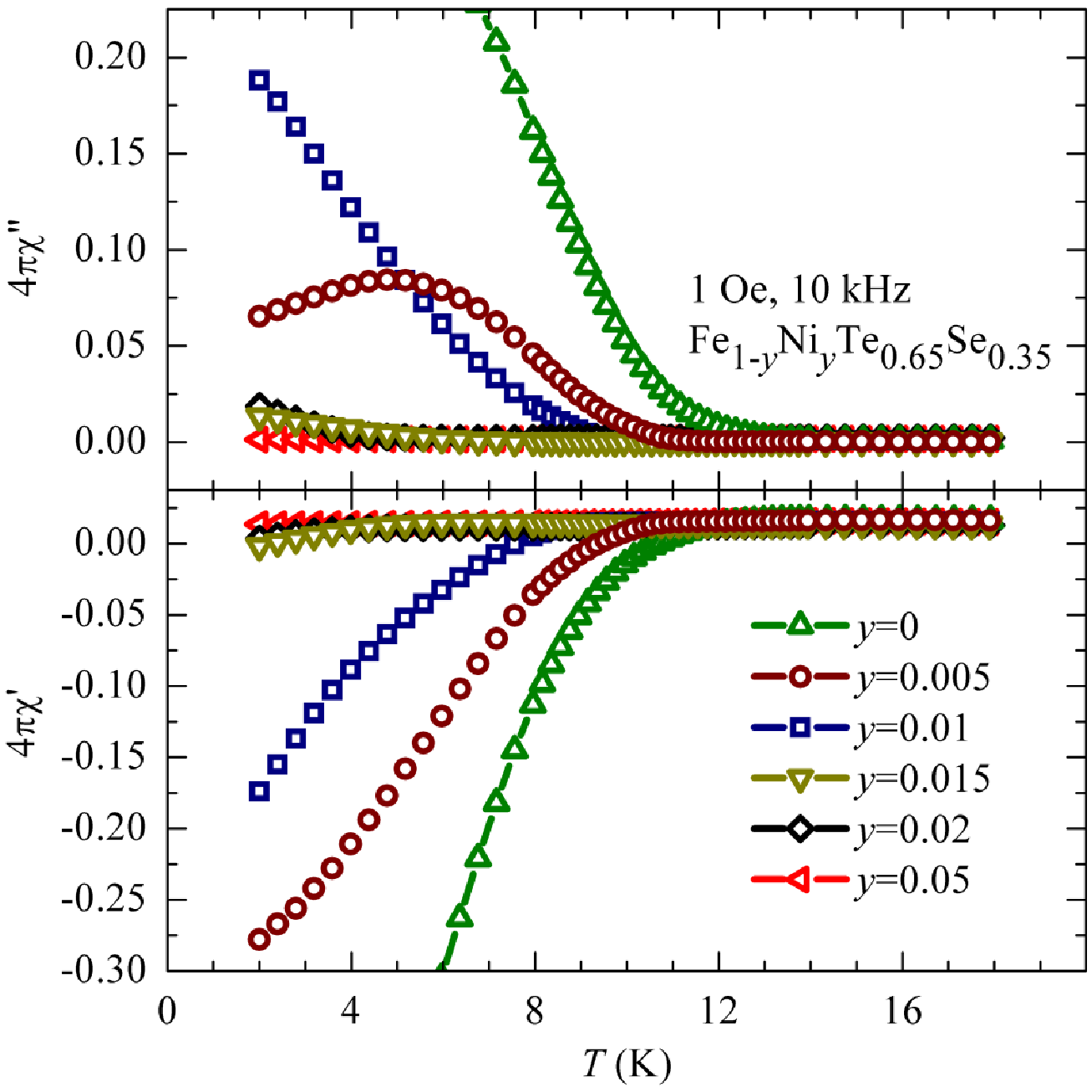

**Figure 10.** Temperature dependence of real part (lower panel) and imaginary part (upper panel) of AC susceptibility for high crystallographic quality $Fe_{1-y}Ni_yTe_{0.65}Se_{0.35}$ ($0.005 < y < 0.05$) single crystals (see, table 2 for Δω) compared with that one for similar crystallographic quality $FeTe_{0.65}Se_{0.35}$ crystals (Δω = 1.67 arc min) measured in 1 Oe of AC field with 10 kHz.

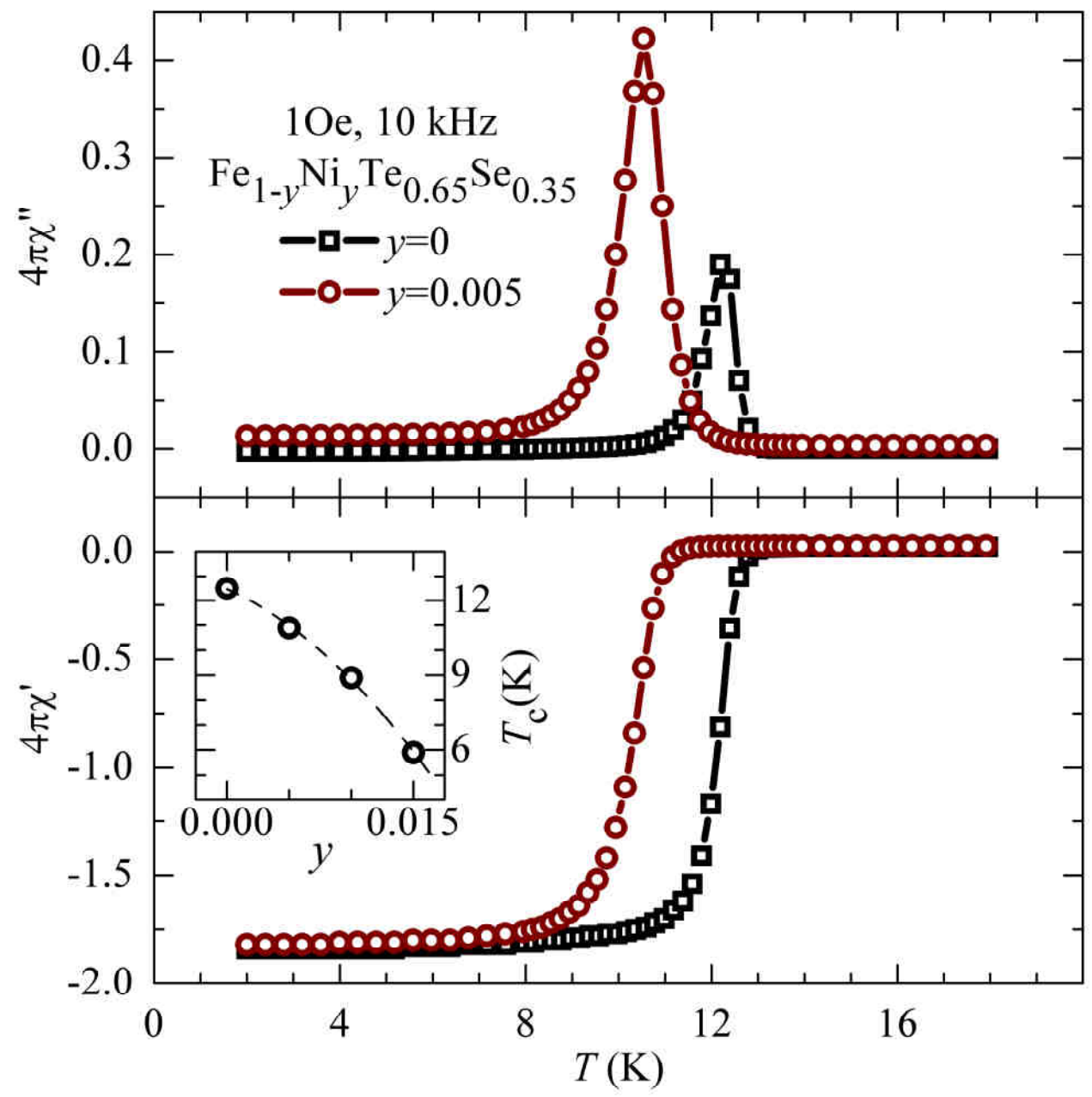

**Figure 11.** Temperature dependence of real part (lower panel) and imaginary part (upper panel) of AC susceptibility for $Fe_{0.995}Ni_{0.005}Te_{0.65}Se_{0.35}$ ($\Delta\omega$ = 5.70 arc min) and $FeTe_{0.65}Se_{0.35}$ ($\Delta\omega$ = 6.00 arc min) single crystals measured in 1 Oe of AC field with 10 kHz - comparison of single crystals with similar crystallographic quality. Inset shows critical temperature dependence on Ni content.

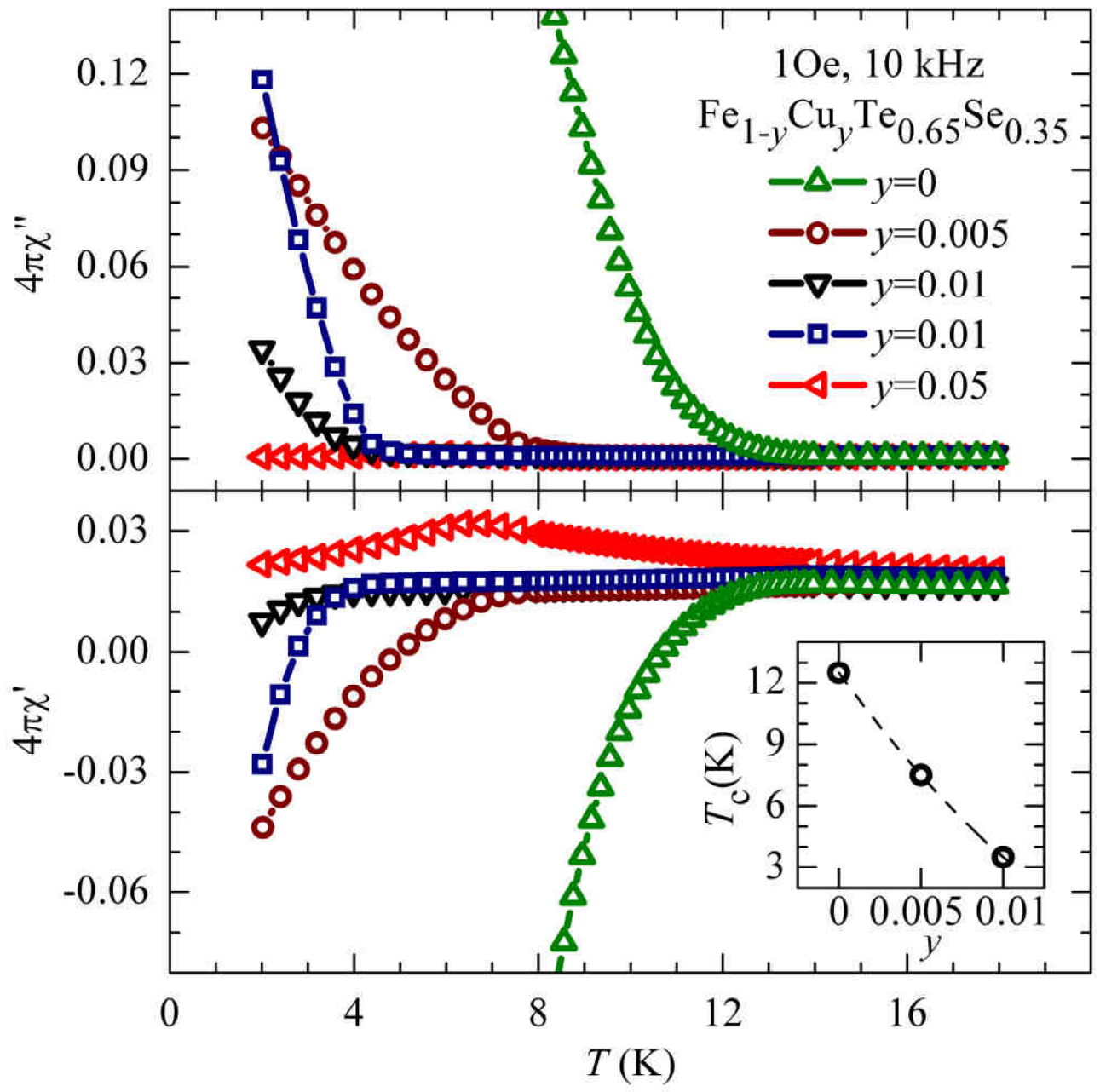

**Figure 12.** Temperature dependence of real (lower panel) and imaginary part (upper panel) of AC susceptibility for $Fe_{1-y}Cu_yTe_{0.65}Se_{0.35}$ ($0.005 < y < 0.1$) single crystals (see, table 2 for $\Delta\omega$) compared with that one for similar crystallographic quality $FeTe_{0.65}Se_{0.35}$ crystals ($\Delta\omega$ = 1.67 arc min) measured in 1 Oe of AC field with 10 kHz. Inset shows critical temperature dependence on Cu content.

## 4. Conclusions

Superconducting single crystal of pure $FeTe_{1-x}Se_x$ ($x = 0.3 - 0.55$) and $FeTe_{0.65}Se_{0.35}$ doped with Co, Ni, Cu, Mn, Zn, Mo, Cd, In, Pb, Hg, V, Ga, Mg, Al, Ti, Cr, Sr or Nd into Fe ions site have been grown using Bridgman's method. It has been found that the sharpness of transition to the superconducting state in $FeTe_{1-x}Se_x$ is evidently inversely correlated with crystallographic quality of the crystals. Crystals with higher value of FWHM exhibit narrower transition to superconducting state. As $\Delta\omega$ value decreases, indicating a decrease of defect concentration, the width of the transition increases.

It has been found that, among examined substitutions of Fe site by Co, Ni, Cu, Mn, Zn, Mo, Cd, In, Pb, Hg, V, Ga, Mg, Al, Ti, Cr, Sr and Nd, only Co, Ni and Cu ions are incorporated at Fe site into host lattice of $FeTe_{0.65}Se_{0.35}$ single crystals. All of the single crystals substituted with Co over 5 at% and with Ni over 2 at% as well as with Cu over 1.5 at%, are not superconducting. The crystals of $Fe_{0.955}Co_{0.045}Te_{0.65}Se_{0.35}$, $Fe_{0.985}Ni_{0.015}Te_{0.65}Se_{0.35}$ and $Fe_{0.99}Cu_{0.01}Te_{0.65}Se_{0.35}$ and those with lower concentration of dopant exhibit superconductivity. The rest of the studied dopants do not incorporate to the host lattice of $FeTe_{0.65}Se_{0.35}$ single crystals. Inclusions, formed with elements not incorporated to the matrix, change chemical composition of the crystals, leading to the changes in Se/Te ratio, what has an impact on the value of critical temperature. The sharpness of transition to the superconducting state is strongly correlated with crystallographic quality of the crystals, similarly as in the case of undoped crystals.

Small disorder introduced into magnetic sublattice, by partial replacement of Fe ions with nonmagnetic ions of Cu or with magnetic ions of Ni or Co with spin value different than that of Fe ion, strongly suppresses superconductivity in $FeTe_{1-x}Se_x$ system. It means that even if superconductivity can appear in the system containing magnetic ions it will not survive if the disorder in magnetic ions sublattice is introduced, most likely because of magnetic scattering of Cooper pairs. This is an indication of the s-wave pairing in superconducting Fe-Se-Te system.

## Acknowledgements

This work was supported by the EC through the FunDMS Advanced Grant of the European Research Council (FP7 ”Ideas”). We thank Tomasz Dietl for suggesting this research and valuable discussions.

**Table 1.** Summary of the chemical composition and structural parameters for selected single crystals of $FeTe_{1-x}Se_x$.

| Starting composition | Composition by EDX (±0.02) | *a* (Å) | *c* (Å) | *V* (Å$^3$) | Δω (min) |
|---|---|---|---|---|---|
| $FeTe_{0.5}Se_{0.5}$ | $Fe_{0.98}Te_{0.57}Se_{0.43}$ | 3.7992 | 6.0560 | 87.412 | 16.65 |
| $FeTe_{0.65}Se_{0.35}$ | $Fe_{0.99}Te_{0.66}Se_{0.34}$ | 3.8036 | 6.0921 | 88.137 | 6.00 |
| $FeTe_{0.65}Se_{0.35}$ | $Fe_{0.99}Te_{0.66}Se_{0.34}$ | 3.7985 | 6.0918 | 87.896 | 3.28 |
| $FeTe_{0.65}Se_{0.35}$ | $Fe_{1.03}Te_{0.65}Se_{0.35}$ | 3.8012 | 6.0874 | 87.958 | 2.52 |
| $FeTe_{0.65}Se_{0.35}$ | $Fe_{0.99}Te_{0.67}Se_{0.33}$ | 3.8020 | 6.0937 | 88.086 | 1.67 |
| $FeTe_{0.7}Se_{0.3}$ | $Fe_{0.98}Te_{0.7}Se_{0.3}$ | 3.7995 | 6.1265 | 88.443 | 2.7 |

**Table 2.** Summary of the chemical composition, structural parameters, and critical temperature for selected single crystals of $Fe_{1-y}TM_yTe_{0.65}Se_{0.35}$ ($TM$ = Co, Ni, Cu, $0.005 < y < 0.2$). The critical temperature $T_c^{onset}$ (10% criterion) was determined from the measurements of AC magnetic susceptibility (field amplitude 1 Oe, frequency 10 kHz) performed with Physical Property Measurement System (PPMS). Symbol <2? in the column with $T_c^{onset}$ means that no trace of superconductivity was detected in the measurements for $T > 2$ K.

| at% of dopant | Composition by EDX (±0.02) | $a$ (Å) | $c$ (Å) | Δω (min) | $T_c^{onset}$ (K) |
|---|---|---|---|---|---|
| 0.5 at% of Co | $Fe_{1.02}Co_{0.01}Te_{0.66}Se_{0.34}$ | 3.7992 | 6.0902 | 1.92 | 11.9 |
| 1 at% of Co | $Fe_{1.04}Co_{0.01}Te_{0.67}Se_{0.33}$ | 3.8034 | 6.0868 | 1.25 | 11.4 |
| 1.5 at% of Co | $Fe_{1.01}Co_{0.017}Te_{0.66}Se_{0.34}$ | 3.8027 | 6.0864 | 1.85 | 10.9 |
| 2 at% of Co | $Fe_{0.98}Co_{0.02}Te_{0.64}Se_{0.36}$ | 3.8021 | 6.0810 | 2.75 | 10.2 |
| 2.5 at% of Co | $Fe_{0.98}Co_{0.03}Te_{0.65}Se_{0.35}$ | 3.8007 | 6.0834 | 1.82 | 9.5 |
| 3 at% of Co | $Fe_{0.97}Co_{0.03}Te_{0.65}Se_{0.35}$ | 3.8044 | 6.0828 | 2.15 | 8.7 |
| 3.5 at% of Co | $Fe_{0.99}Co_{0.04}Te_{0.65}Se_{0.35}$ | 3.8005 | 6.0817 | 2.03 | 8.2 |
| 4 at% of Co | $Fe_{1.01}Co_{0.04}Te_{0.68}Se_{0.32}$ | 3.7994 | 6.0794 | 1.82 | 7.2 |
| 4.5 at% of Co | $Fe_{0.96}Co_{0.05}Te_{0.65}Se_{0.35}$ | 3.8017 | 6.0846 | 1.85 | ~5.9 |
| 5 at% of Co | $FeCo_{0.06}Te_{0.67}Se_{0.33}$ | 3.8025 | 6.0792 | 1.82 | <2? |
| 10 at% of Co | $Fe_{0.93}Co_{0.10}Te_{0.66}Se_{0.34}$ | 3.7996 | 6.0662 | 2.85 | <2? |
| 15 at% of Co | $Fe_{0.87}Co_{0.13}Te_{0.69}Se_{0.31}$ | 3.7986 | 6.0545 | 2.03 | <2? |
| 20 at% of Co | $Fe_{0.90}Co_{0.16}Te_{0.66}Se_{0.34}$ | 3.7990 | beginning 6.0344<br>end 6.0493 | 3.00<br>6.00 | <2? |
| 0.5 at% of Ni | $Fe_{0.94}Ni_{0.007}Te_{0.66}Se_{0.34}$ | 3.8046 | 6.0849 | 5.70 | 11.7 |
| 0.5 at% of Ni | $Fe_{1.04}Ni_{0.008}Te_{0.7}Se_{0.3}$ | 3.8034 | 6.0878 | 2.45 | 10.9 |
| 1 at% of Ni | $Fe_{1.01}Ni_{0.01}Te_{0.66}Se_{0.34}$ | 3.8014 | 6.0881 | 3.38 | 8.9 |
| 1.5 at% of Ni | $Fe_{0.98}Ni_{0.015}Te_{0.66}Se_{0.34}$ | 3.8006 | 6.0875 | 1.97 | ~5.9 |
| 2 at% of Ni | $Fe_{1.02}Ni_{0.02}Te_{0.66}Se_{0.34}$ | 3.8002 | 6.0868 | 1.62 | <2? |
| 5 at% of Ni | $Fe_{0.98}Ni_{0.06}Te_{0.66}Se_{0.34}$ | 3.8005 | 6.0815 | 1.25 | <2? |
| 10 at% of Ni | $Fe_{0.93}Ni_{0.11}Te_{0.64}Se_{0.36}$ | 3.7997 | 6.0555 | 2.55 | <2? |
| 20 at% of Ni | $Fe_{0.87}Ni_{0.21}Te_{0.64}Se_{0.36}$ | 3.8015 | beginning 6.0288<br>end 6.0275 | 1.73<br>3.45 | <2? |
| 0.5 at% of Cu | $Fe_{1.03}Cu_{0.007}Te_{0.66}Se_{0.34}$ | 3.8010 | 6.0911 | 1.70 | 7.6 |
| 1 at% of Cu | $Fe_{1.01}Cu_{0.01}Te_{0.65}Se_{0.35}$ | 3.8022 | 6.0865 | 1.47 | ~3.5 |
| 1 at% of Cu | $Fe_{1.01}Cu_{0.01}Te_{0.67}Se_{0.33}$ | 3.8026 | 6.0901 | 2.27 | <2? |
| 5 at% of Cu | $Fe_{0.94}Cu_{0.06}Te_{0.65}Se_{0.35}$ | 3.8017 | 6.0682 | 1.92 | <2? |
| 10 at% of Cu | $Fe_{0.89}Cu_{0.13}Te_{0.65}Se_{0.35}$ | 3.8008 | 6.0486 | 1.35 | <2? |
| 20 at% of Cu | $Fe_{0.84}Cu_{0.21}Te_{0.65}Se_{0.35}$ | 3.8003 | beginning 6.0272<br>end 6.0250 | 1.65<br>3.35 | <2? |

**Table 3.** Summary of the chemical composition of matrix and inclusions, structural parameters and critical temperature for selected single crystals of $FeTe_{0.65}Se_{0.35}$ doped with the elements not incorporated into the host lattice.

| at% of dopant | Matrix composition by EDX (±0.02) | Inclusions |
|---|---|---|
| 5 at% of Zn | $Fe_{1.01}Te_{0.69}Se_{0.31}$ | ZnSe |
| 5 at% of Mn | $Fe_{1.09}Te_{0.69}Se_{0.31}$ | $MnSe+(FeMn)_2O_3$ |
| 20 at% of Mn | $Fe_{1.04}Te_{0.78}Se_{0.22}$ | |
| 5 at% of Cd | $Fe_{1.03}Te_{0.66}Se_{0.34}$ | $CdTe_{1-x}Se_x$ |
| 2 at% of In | $FeTe_{0.65}Se_{0.35}$ | In-Te |
| 5 at% of Mo | $Fe_{1.15}Te_{0.77}Se_{0.32}$ | $Mo(Te_{0.3}Se_{0.7})_2+Mo(Te_{0.4}Se_{0.6})_2$ |
| 2 at% of Pb | $Fe_{1.05}Te_{0.7}Se_{0.3}$ | Fe-Pb-Te |
| 2 at% of Hg | $Fe_{0.99}Te_{0.67}Se_{0.33}$ | Fe-Hg-Te |
| 5 at% of V | $Fe_{1.07}Te_{0.68}Se_{0.32}$ | $V_{0.65}Fe_{0.35}$ |
| 1 at% of Ga | $Fe_{0.97}Te_{0.66}Se_{0.34}$ | $Fe_{0.4}Ga_{0.6}$:O |
| 5 at% of Mg | $Fe_{1.03}Te_{0.66}Se_{0.34}$ | $Mg_{0.55}Si_{0.45}$:Na:K:O |
| 1 at% of Al | $Fe_{1.01}Te_{0.66}Se_{0.34}$ | precipitations with dopant elements on the top of the ingot |
| 5 at% of Ti | $Fe_{1.05}Te_{0.67}Se_{0.33}$ | |
| 5 at% of Cr | $Fe_{1.05}Te_{0.66}Se_{0.34}$ | |
| 1 at% of Sr | $Fe_{1.01}Te_{0.66}Se_{0.34}$ | |
| 1 at% of Nd | $Fe_{1.04}Te_{0.67}Se_{0.33}$ | |